\documentclass[journal,12pt,draftclsnofoot,onecolumn,comsoc]{IEEEtran}
\usepackage[T1]{fontenc}

\ifCLASSINFOpdf
\else
\fi
\usepackage{amsmath}
\usepackage{txfonts}
\newcommand\Mark[1]{\textsuperscript#1}
\def\be{\begin{equation}}
\def\ee{\end{equation}}

\newcommand{\argmax[1]}{\mathrm{arg} \underset{#1} {\mathrm{max}}\,}
\newcommand{\Pp}{\mathcal{P}}

\newtheorem{mythe}{Theorem}
\newtheorem{mylem}{Lemma}
\newtheorem{mypro}{Proposition}

\usepackage{algorithm}
\usepackage{algorithmic}
\usepackage{graphicx}
\usepackage{caption}

\begin{document}
%
\title{Network Resource Sharing Games with Instantaneous Reciprocity}
%
%
%


\author{\IEEEauthorblockN{Sofonias Hailu\Mark{1}, Ragnar Freij-Hollanti\Mark{1}, Alexis A. Dowhuszko\Mark{2}, and Olav Tirkkonen\Mark{1}}\\
\Mark{1}Department of Communications and Networking, Aalto University, P.O. Box 13000, FI-00076 Aalto, Finland\\
\Mark{2}Centre Tecnol\`{o}gic de les Telecomunicacions de Catalunya (CTTC), Barcelona, Spain\\
E-mail: \{sofonias.hailu, ragnar.freij, olav.tirkkonen\}@aalto.fi,  alexis.dowhuszko@cttc.es}

\maketitle

\begin{abstract}
We propose a generic strategic network resource sharing game between a set of players representing operators. The players negotiate which sets of players share given resources, serving users with varying sensitivity to interference. We prove that the proposed game has a Nash equilibrium, to which a greedily played game converges. Furthermore,  simulation results show that, when applied to inter-operator spectrum sharing in small-cell indoor office environment, the convergence is fast and there is a significant performance improvement for the operators when compared to the default resource usage configuration.
\end{abstract}


%
\IEEEpeerreviewmaketitle

\section{Introduction}
Game theory has been widely used in analyzing and designing wireless
network protocols. Often, game theoretical principles have been used
as guiding light when striving for distributed solutions of NP-hard
optimization problems. The idea is that if in a given networking
situation, a Nash Equilibrium (NE) of a strategic game played by
transmitters and receivers in the network is reasonably close to a
Pareto optimal operation point, simple distributed implementations can
be found. This kind of solutions have been searched for mostly related
to the physical and Medium Access Control layers of cellular and
ad Hoc networks.

On the physical layer, distributed power control and power allocation
based on strategic games have been widely studied.
In~\cite{MacKenzie2001}, power control cellular systems with Code
Division Multiple Access (CDMA) were addressed. Multichannel power
control between transmitter--receiver (Tx-Rx) pairs based on iterative
water-filling game was addressed in~\cite{Yu2002}.
In the setting of selfish Tx-Rx pairs operating in unlicensed bands,
it was observed that one-shot games of players with full freedom to
allocate power leads to socially suboptimal power allocations, where
power is distributed over the full bandwidth~\cite{Etkin2007}.
Repeated game approaches to cure this were considered
in~\cite{Etkin2007,Wu2009,Bennis2009}. In these, first an agreement is
reached about power allocation over spectral resources, either a
Pareto efficient point~\cite{Etkin2007}, orthogonal
allocation~\cite{Wu2009}, or a social optimum~\cite{Bennis2009}. The
agreed resource allocation is maintained with a grim
trigger~\cite{Etkin2007,Bennis2009}, or a finite period punishment
strategy~\cite{Wu2009}. An alternative to the repeated game solution
would be a cooperative game approach. In \cite{Suris2007}, Nash
Bargaining is used to agree on a fair and efficient allocation of
spectrum.

Recently, power allocation in frequency selective fading channels was
reconsidered in a network of strategic Tx-Rx
pairs~\cite{Bistritz2015}. In this case, the pre-agreement states that
players use their $M$ best channels, so that the received SINR is the
same on all used channels. This strategy is shown to lead to Pure NEs
which are asymptotically socially optimal when the number of players
approaches infinity.

%
%

In a higher layer view of resources allocation, one is interested not
in the power allocation per se, but on which resources are used by
which players. Potential game approaches~\cite{Neel2006,Yamamoto2015}
have been successfully used to solve many discrete resource allocation
problems, in situations where the players utility functions are
aligned with a global potential function. In~\cite{Neel2006}, spectrum
was used as a discrete resource, which is either used or not used, and
each player is constrained by hardware to choose only one channel.

In more involved scenarios, cooperative spectrum sharing games have
been played between cellular network operators~\cite{Kamal2009,
  Si2010}. In these cooperative approaches, there is a component of
spectrum pricing involved, which penalizes increased spectrum usage.

The problem of spectrum sharing between operators~\cite{Kamal2009,
  Si2010} differs from, e.g., physical layer power allocation problems
addressed
in~\cite{Yu2002,Etkin2007,Wu2009,Bennis2009,Suris2007,Bistritz2015} by
the status of the players of the game. In physical layer settings,
such as power allocation, game theory acts as an inspiration to
designing distributed algorithms. These algorithms would be implemented
in hardware, and typically there would be a standard governing the
implementation. Conformance tests would then apply to the hardware,
and the hardware entities would have no independent rationality
allowing generic change of strategy. Mechanism design~\cite{Hurwicz2006} would then lead to hardware implementation. A good example of
this is~\cite{Bistritz2015}, where a fully distributed, almost
socially optimal algorithm based on a mechanism obeyed by all resource
allocation players was presented. 

%
In the multioperator spectrum sharing problem, the players are instead
entities with full freedom of action. Decisions on which set of
carriers is used for communication, are taken by truly economic actors,
or by software implementations governed by such actors. In such
settings, mechanism design would take the form of designing
protocols that enable socially beneficial behavior, which are enforced either by law, or by legally binding agreements. 

 In~\cite{Hailu2014,Singh2015}, we
have investigated mechanisms defined by coordination protocols
determining allowed sets of actions of players participating in
network resource allocation games, where there is interference between
the networks, when they use the same resource. A mechanism based on
{\it
  instantaneous reciprocity} was discussed in~\cite{Hailu2014}.
Scenarios of {\it mutual renting}, where each player has a private
resource, and a {\it resource pool}, where each player has equal right
to access resources, were addressed. It was shown that a protocol
where resources are divided by the individual players and the set of
all players has a dominant strategy Nash equilibrium. Players are
willing to sacrifice some of their right to use resources, if all
others do the same. This concept differs from reciprocal altruism,
studied in~\cite{Axelrod1981}. According to~\cite{Axelrod1981},
reciprocal altruism emerges in indefinitely repeated strategic games.
Strategies incentivizing socially optimal behavior based on
reciprocation and forgiveness can be found. 
In~\cite{Hailu2014},  reciprocity is instantaneous and dictated by the
mechanism.  

In this paper we study a general strategic resource allocation game
between $N$ competing parties. There is a network resource utilization
pattern, determining which players use which resource. There is
interaction between the players when they use the same resources---we
assume that the utility function of individual players are concave in
these patterns. The mechanism enforced by a coordination protocol is
based on instantaneous reciprocity, but differing
from~\cite{Hailu2014}, {\it any subset of players} may reach
agreement about reciprocal resources usage, and {\it any number of
  such subsets} may have simultaneous agreements. We thus have a
reciprocal resource partitioning game in the {\it the power set of the
  users}. The motivation of the problem setting is that the players
would be serving multiple users, and accordingly, there are varying
degrees of conflict between resource usage, depending on the amount of
resources shared with different subsets of other players. This leads
to preferences to play with multiple different subsets.

Despite the fact that the problem of finding subsets of players with
similar interest, the considered 
game is {\it not} a coalition formation game.\footnote{Coalitional
  games (see \cite{Saad2009}) provide a framework for players to join
  forces and reach non-zero-sum outcomes.
} 
A generic player may prefer to be part of multiple coalitions, and
have a preference for the distribution of resources between these
coalitions. The utilities of the players are fully non-transferable,
and the games are $N$-player strategic games. Coalition formation and
bargaining games may be developed based on the strategic games
considered here. This, however, is left for future work.

In the considered powerset resource sharing games, we prove
existence of pure strategy Nash equilibria, and provide a sequential resolution
scheme where best response greedy bidding is proven to converge to a
Nash Equilibrium.

As an example we consider a realization of the game related to
spectrum sharing between operators. We show that if the operators
apply an $\alpha$-fair sum-utility function~\cite{Mo00}, the operator
utility function is concave in the network resource utilization
pattern, and thus the NE results proven for the generic game apply. We
provide simulation results for a game played between $N=4$ operators
in a small-cell indoor office environment.

The paper is organized as follows. Section II describes the system model. The network resource sharing problem is formulated as an $N$-person game in Section III. The existence of equilibrium point is shown in Section IV. In Section V, a sequential $N$-person game is discussed where the players have a greedy strategy and shown to converge. Section VI discusses inter-operator spectrum sharing as an application example. Section VII provides simulation results and analysis. Section VIII draws conclusions.

\section{System model}
\label{sec:SysMod}
There are $N$ players, given by a set $\cal N$, who are negotiating about one unit of a shared resource. The resource is divided into $2^N$ parts, one part $b_S$ for each subset $S \subset {\cal N}$. We say that the fraction $b_S$ is \emph{allocated} to the subset $\cal S$. The resource $b_S$ is non-orthogonally shared by the players who are a member of the subset $\cal S$. We have
\be
 \sum_{S \subset {\cal N}} b_S = 1~.
\ee
Under the natural assumption that no part of the resource is left completely unused, we have $b_{\emptyset} = 0$.

The default usage pattern of the shared resource is determined, for example, by a regulatory body or by previous negotiations. The two most interesting cases are the mutual renting game (MRG), where 
\be
b_S^0 = \begin{cases}
1/N &  \quad \text{if } |S| = 1 \\
0 & \quad \text{otherwise}
\end{cases}
\label{eq:Default_MRG}
\ee
for all $S \subset {\cal N}$, meaning that all resources are private by default, and the resource pool game (RPG), where
\be
b_S^0 = \begin{cases}
1 &  \quad \text{if } |S| = N \\
0 & \quad \text{otherwise}
\end{cases}
\label{eq:Default_RPG}
\ee
for all $S \subset {\cal N}$., meaning that all resources are unlicensed by default or all the players has a right to use the shared resource. In MRG, we assume without loss of generality that all players have the same amount of resource to start with.

For a given resource usage pattern $\mathbf{b} \in \{\mathbb{R}^+\}^{2^N}$, the players get a utility of $g_n(\mathbf{b})$ for all $n \in \cal N$. We assume that each function $g_n$ is strictly concave and differentiable in $\mathbf{b}$. We will verify in Section~\ref{sec:AppInt} that this assumption is valid and natural in a wide variety of network problems, where each player is serving several users, and her payoff is an alpha-fair summation of the experience of these users. There are precisely $2^{N-1}$ subsets $S \subset \cal N$ such that $n\in S$, and we denote the collection of these sets by ${\cal P}_n$.
 The utility $g_n(\mathbf{b})$ is assumed to depend only on $\{b_S : S\in {\cal P}_n\}$ changes only when the value of $b_S$ such that $S \in {\cal P}_n$ varies. Consider two resource utilization patterns $\mathbf{b}_1, \mathbf{b}_2$ and two subsets $S \in {\cal P}_n$ and ${\cal {\tilde{S}}} \subset {\cal N}$. Let $b_{1\cal T} = b^0_{\cal T}$ for all ${\cal T} \neq \cal S$, $b_{1\cal S} = b^0_S - \epsilon$, $b_{2\cal T} = b^0_{\cal T}$ for all ${\cal T} \neq {\cal {\tilde{S}}}$, $b_{2{\cal {\tilde{S}}}} = b^0_{\cal {\tilde{S}}} - \epsilon$, and $\epsilon > 0$. We have the following order relationship
\be
 \tilde{ S} \subsetneq { S}~~\Rightarrow~~ g_n(\mathbf{b}_2) < g_n(\mathbf{b}_1)
\label{eq:OrdRel}
\ee
In terms of network games we interpret this as follows: The players' utility functions are increasing in the experience of their users, and this experience increases the less interference is experienced from other players.
Note that this holds also if
$\tilde{\cal S} \not\in {\cal P}_n$, as then $g_n(\mathbf{b}_2) = g_n(\mathbf{b}^0)$.

The players negotiate to determine a resource usage pattern $\mathbf{b}$ that is valid for a given period called resource sharing period. The objective of the players is to maximize their utility function $g_n$. In Section \ref{sec:ProFor}, we define a game that serves as a formalism for these negotiations.

\section{Problem formulation as $N$-person game}
\label{sec:ProFor}
The resource sharing problem is formulated as an $N$-person game. The game is denoted as $\Gamma = \{ {\cal N}, \{ \mathbf{a}_n\}_{n \in \cal N},\{ \Phi_n(\mathbf{a})_{n \in \cal N} \}$. 
The strategy of a player, $\mathbf{a}_n \in \{\mathbb{R} \cup *\}^{2^N}$ will be interpreted as the values of $b_S$ that are prefered by player $n$. Here, $a_{nS}=*$ is interpreted as $n$ not having an opinion about the resources assigned to set $S$, which will be applicable when $n\not\in S$. A joint strategy of all players will therefore be a matrix in $\{\mathbb{R} \cup *\}^{2^N\cdot N}$, where half the entries take the void value $*$. The payoff of a player, $\Phi_n(\mathbf{a}) \in \mathbb{R}^+$, is given as the utility of the player with the agreed resource usage pattern. The game is determined by two things: an instantaneous reciprocity and the \emph{a priori} rule $\mathbf{a}\mapsto\mathbf{b}$ to determine the outcome of the negotiations.
Hence, after defining an outcome $\mathbf{b}=\Upsilon(\mathbf{a})\in \{\mathbf{R}\}^{2^N}$ as a function of the strategies (henceforth \emph{bids}) $\mathbf{a}$, we can define the payoff function $\Phi_n$ as \be \label{eq:payoff} \Phi_n(\mathbf{a})=g_n(\Upsilon(\mathbf{a})).\ee

\subsection{Instantaneous reciprocity}
The strategy of a player in a given resource sharing period does not depend on the outcome of the games in the past or future spectrum sharing periods. Since the players are assumed to be selfish, there has to be an instantaneous reciprocity. In particular, we have the constraint
\be
 \sum_{S\in{\cal P}_n} \frac{b_S}{|S|} = \frac1N,
\label{eq:InsRes}
\ee
that ensures that all players give and take the same amount of favor.
Here, in MRG, a favor is to give right to use on a resource. In RPG, a favor is to remain silent on a resource. A favor given by a player is divided among the players using it. 

We hence define the set $\mathcal{F}$ of feasible sharing patters by 
\be
\mathcal{F}=\left\{b\in\mathbb{R}^{2^N} {\huge\vert} \begin{split}  0 & \leq b_S & \hbox{ for all } S\subseteq\mathcal{N}\\ \sum_{S\in\mathcal{P}_n} \frac{b_S}{|S|} &= \frac{1}{N} & \hbox{ for all } n\in \mathcal{N}\end{split}\right\}
\ee Hence, $\mathcal{F}$ is a convex polytope defined by $2^N$ inequalities and $N$ equalities.
Note that the distinction between MRG and RPG is not observed in the set of feasible resource sharing patterns, but only in the default sharing pattern, and hence in the resolution rule, which is defined in Section~\ref{ssec:rule}. 

\subsection{Strategy space} 
Let ${\cal A}_n$ be the strategy space of player $n$, meaning the set of allowed bids. We define ${\cal A}$ as the subspace ${\cal A}_n \subset \{\mathbb{R} \cup *\}^{2^N}$ given by the constraints
\be
\begin{aligned}
\sum_{S\in{\cal P}_n} \frac{a_S}{|S|} &= \frac1N , \\
a_{n\cal S} &= * & \forall n \notin S, S \subset {\cal N}, \\
a_{n\cal S} &\geq 0 & \forall n \in S, S \subset {\cal N}
\end{aligned}
\label{eq:StrSpa}
\ee
which is a compact convex set. This means that $\mathbf{a}_n$ has to respect the instantaneous reciprocity constraint~\eqref{eq:InsRes}. Moreover, if $n\not\in S$, then we require $a_{nS}$ to take the void value $*$, which should be interpreted as not having any prefered value of $b_S$. This is natural to require, to not allow players to obstruct an agreement that does not affect her payoff, between a set of players in which she is not a member.

\subsection{An \textit{a-priori} resolution rule}\label{ssec:rule}
The game has an \textit{a-priori} agreed rule to resolve the strategies of the players. Denote the resolution rule by \[\Upsilon:\left(\mathbb{R}\cup *\right)^{N\cdot 2^N} \rightarrow \mathbb{R}^{2^N},\] mapping the joint strategy of the players $\{ \mathbf{a}_n\}_{n \in \cal N}$ to the resource allocation $\mathbf{b} \in \mathbb{R}^{2^N}$. The resolution rule $\Upsilon$ will also depend on a default resource distribution $\mathbf{b}^0\in \mathcal{F}$, as described in Section~\ref{sec:SysMod}. 
We also propose a sequential game, where the outcome of one round will be the default distribution of the next. This is only one of the reasons why we allow arbitrary $\mathbf{b}^0\in\mathcal{F}$.

For a bid $\mathbf{a}_n\in\left(\mathbb{R}\cup *\right)^{2^N}$, define the set \[I_n=\prod_{S\in\Pp_n} [\min(\mathbf{a}_{n S}, \mathbf{b}^0_S) , \max(\mathbf{a}_{n S}, \mathbf{b}^0_S)]\times\prod_{S\not\in \Pp_n}\mathbb{R}.\] Requiring the outcome $\{b_S :n\in S\}$ to remain within $I_n$, we guarantee that no player has to change her default usage pattern more than she is willing to according to her bid.
We propose the following resolution rule:
\be 
\begin{aligned}
\Upsilon(\{ \mathbf{a}_n\}_{n \in \cal N}) = \, & \argmax[\mathbf{b}] 
& \sum_{S \subset {\cal N}} |\mathbf{b}_S - \mathbf{b}^0_S|\\
& \text{subject to}
&  \mathbf{b} \in {\cal{F}} \\
& & \mathbf{b}\in I_n \hbox{ for all } n\in \mathcal{N}.
\end{aligned}
\label{eq:ResRul}
\ee
Observe that, while in general the maximum might not be unique, it will be almost surely, assuming the bids are drawn from a continuous probability distribution. This will in turn happen if the players play optimally, and the players' utility functions $g_n$ are generic enough. We may also observe that, while it is in general computationally unfeasible to find the largest vector in a given set, the resolution rule can be evaluated efficiently, thanks to the following lemma.
\begin{mythe}\label{thm:linear}
The resolution rule $\Upsilon$ is the solution of a linear optimization problem.
\end{mythe}
\begin{IEEEproof}The proof is easy but technical, and is postponed to Appendix~A.
\end{IEEEproof}

\section{Existence of equilibrium point}
\label{sec:ExiEqu}
We will begin by stating and proving a standard game theoretic lemma in a version that suits our settings. For game-theoretic terminology, we refer to~\cite{gamebook}, and for topological notions we refer to~\cite{munkres}.
\begin{mylem}
Let $(\mathcal{N},\mathcal{A},\Phi)$ be an $N$-player game that satisfies the following three criteria: 
\begin{enumerate}  
\item The payoff function $\Phi$ is upper semicontinuous.
\item The joint strategy space $\mathcal{A}$ is convex and compact. 
\item For each player $n\in \mathcal{N}$, and all joint strategies $\mathbf{a}_{-n}=\{a_{iS} : i\neq n\}$, the set \[\gamma_n (\mathbf{a}_{-n})=\argmax[\mathbf{a}_n] \Phi(\mathbf{a})\] is convex.
\end{enumerate}
Then $(\mathcal{N},\mathcal{A},\Phi)$ has a Nash equilibrium point.
\label{nashcriterion}
\end{mylem}
\begin{IEEEproof}
Let $P(\mathcal{A})$ be the set of all subspaces of $\mathcal{A}$. Consider the set valued map $\nabla: \mathcal{A}\to P(\mathcal{A})$, given by $\nabla(\mathbf{a})=\prod_n\gamma_n (\mathbf{a}_{-n})$. By upper semicontinuity of $\Phi$, $\gamma_n$ is closed, and by assumption (3) it is also convex. As $\nabla(\mathbf{a})$ is a product of finitely many compact convex sets, it is itself compact and convex. It also follows from upper semicontinuity of $\Phi$ that the $\nabla$ is upper hemicontinuous. Now by Kakutani's theorem~\cite{Kakutani41}, any function $X\to P(X)$ on a compact domain $X$, that is upper-hemicontinuous and takes closed and convex values, has a fixed point. Hence, there exists $\mathbf{a}\in \mathcal{A}$ such that $\mathbf{a}\in \nabla(\mathbf{a})=\prod_n\gamma_n (\mathbf{a}_{-n})$. This means that $\mathbf{a}_n\in \gamma_n (\mathbf{a}_{-n})$ for every $n$, which in turn means that no player can improve her outcome by changing her bid. Hence, $\mathbf{a}$ is a Nash equilibrium.\end{IEEEproof}

We now show that~\ref{nashcriterion} applies to the game described in Section~\ref{sec:ProFor}. 
\begin{mythe}
The game  \[(\mathcal{N},\mathcal{A}=\prod_{n\in \mathcal{N}}\mathcal{A}_n,\Phi=g\circ\Upsilon)\] described in Section~\ref{sec:ProFor} has a Nash equilibrium.
\label{ProOne}
\end{mythe}
\begin{IEEEproof}
The payoff function $\Phi$ is upper semicontinuous by definition.\footnote{To be precise, this is only true when we have defined a tiebreak in the points where the function $\Upsilon$ is not uniquely defined. This is a mere technicality as such situations will almost never occur in practice. We can extend the function $\Phi$ to these points, upper semicontinuously for all players.} The joint strategy space $\mathcal{A}=\prod_{n\in \mathcal{N}}\mathcal{A}_n$ is a product of finitely many compact spaces, and is therefore compact in its own right. To apply Theorem~\ref{nashcriterion}, we need to consider the sets \[\gamma_n (\mathbf{a}')=\argmax[\mathbf{a}_n] \Phi(\mathbf{a})\] for fixed $\mathbf{a}'$ and $n$. This is the set of points $\mathbf{a}_n$ such that $\Upsilon(\{\mathbf{a_i}\}_{i\in\mathcal{N}})$ maximizes $g_n$ over all \[\mathbf{b}\in \mathcal{F}\cap(\cap_{i\in\mathcal{N}-\{n\}}I_i).\] 

By strict concavity of $g_n$, it is maximized in a unique point $\mathbf{c}\in \mathcal{F}_n$, where \[\mathcal{F}_n=\mathcal{F}\cap(\cap_{i\in\mathcal{N}-\{n\}}I_i)\] Now $\gamma_n$ can be written as \[\gamma_n=\{\mathbf{a}_n\in \mathcal{A}_n | \argmax[\mathbf{b}\in \mathcal{F}_n\cap I_n]\|\mathbf{b} - \mathbf{b}^0\|_1 = \mathbf{c}\}.\] But this set is given by the inequalities \[\begin{aligned} a_{nS} &\geq c_{S} & \hbox{if } c_{S}=\min\{\mathbf{a}_{iS} : i\neq n\}\geq \mathbf{b}_S^0\\
&&\hbox{or } c_{S}=\mathbf{b}_S^0 >\max\{\mathbf{a}_{iS} : i\neq n\}, \\
a_{nS} &\leq c_{S} & \hbox{if } c_{S}=\max\{\mathbf{a}_{iS} : i\neq n\}\leq \mathbf{b}_S^0\\
&&\hbox{or } c_{S}=\mathbf{b}_S^0 <\min\{\mathbf{a}_{iS} : i\neq n\}, \\
a_{nS} &=c_{S} & \hbox{if } \max\{\mathbf{a}_{iS} : i\neq n\}< c_S < \mathbf{b}_S^0\\
&&\hbox{or } \mathbf{b}_S^0 <c_S < \min\{\mathbf{a}_{iS} : i\neq n\}.
\end{aligned}\]
As this set is clearly convex, Lemma~\ref{nashcriterion} can be applied, so the game has a Nash equilibrium.
\end{IEEEproof}

\section{Sequential $N$-person game}
\label{sec:SeqNpe}
In section~\ref{sec:ExiEqu}, the $N$-person game is shown to have a Nash equilibrium point. However, the Nash equilibrium point is not unique unless we have $\mathbf{a}_n = \mathbf{a}_{n'}$ for all $n,n' \in {\cal N}$. Indeed, as long as any player has bids $a_{nS}\neq b_S, a_{nT}\neq b_T$ that are not equal to the game outcome, then sufficiently small changes can be made to $a_{nS}$ and $a_{nT}$ while respecting the instantaneous resiprocity rule, without changing the outcome of the game. The number of equilibrium points is therefore infinite in the typical case when the Nash equilibrium is not unique. 

Since the players are selfish, they prefer a resource usage pattern that maximizes their utility function. It is therefore natural to define their \emph{greedy strategy} by
\be
\begin{aligned}
\mathbf{a}_n = & \hspace{4mm}\argmax[\mathbf{a}]
& & g_n(\mathbf{a}) \\
& \hspace{4mm} \text{subject to}
& & \mathbf{a} \in {\cal A}_n
\end{aligned}
\label{eq:GreStr}
\ee
for all $n \in \cal N$, with $a_{nS}=*$ if $n\not\in S$. The point $(\mathbf{a}_1,\mathbf{a}_2,\hdots,\mathbf{a}_N)$ is not a Nash-equilibrium point unless $\mathbf{a}_n = \Upsilon(\{ \mathbf{a}_n\}_{n \in \cal N})$ for at least $N-1$ players. The players, thus, may further want to play a $N$-person game with updated $\mathbf{b}^0 \leftarrow \Upsilon(\{ \mathbf{a}_n\}_{n \in \cal N})$. This naturally leads to a sequence of games where the default spectrum utilization pattern is updated at each iteration as summarized in algorithm~\ref{alg:SeqNpe}.

\begin{algorithm}[h]
\caption{Sequential $N$-person game}
\begin{algorithmic}[1]
\STATE INITIALIZATION
\STATE Given $\mathbf{b}^0$
\REPEAT
\STATE Each player $n \in \cal N$ evaluates ${\cal A}_n$ using~\eqref{eq:StrSpa}
\STATE Each player $n \in \cal N$ obtains $\mathbf{a}_n \in {\cal A}_n$ using~\eqref{eq:GreStr}
\STATE $\mathbf{b}^0 \leftarrow \Upsilon(\{ \mathbf{a}_n\}_{n \in \cal N})$
\UNTIL {\text convergence}
\end{algorithmic}
\label{alg:SeqNpe}
\end{algorithm}

\begin{mythe}
The sequential $N$-person game converges. 
\end{mythe}
\begin{IEEEproof}
The strategy of a player remains the same in each iterations, and the new default pattern $\mathbf{b}^0$ is contained in all $I_n^i$. Therefore, the domain of~\eqref{eq:ResRul} forms a decreasing chain, since $(\cap_{n\in \mathcal{N}}I_n^i) \subseteq (\cap_{n\in \mathcal{N}}I_n^{i-1})$. In each iteration, at least one new bid $a_{nS}$ gets satisfied, in the sense that $a_{nS}=b_S$. After this happened, the value of $b_S$ will not change again, since player $n$ would obstruct such a change. Therefore, the sequence domains of~\eqref{eq:ResRul} forms a decreasing sequence of polytopes of strictly decreasing dimension. Thus, the sequence converges to a point, to which therefore also the outcomes $\mathbf{b}$ converge. 
\end{IEEEproof}

If $N = 2$, it can be observed that the sequential game converges in one iteration leading to a one-shot game. This is due to the reason that at the first iteration, the strategy of one of the players is selected, i.e. $\exists n \mid \mathbf{a}_n = \Upsilon(\{ \mathbf{a}^i_n\}_{n \in \cal N})$, leading to the maximum possible payoff improvement for the player. Thus, there is no reason for the player to change the agreed resource utilization pattern.

\begin{mylem}
The sequential $2$-person game converges in one iteration. The strategy of the players is dominant. At least one of the players has a strongly dominant strategy.
\label{lem:DomStr}
\end{mylem} 
\begin{IEEEproof}
Let ${\cal N}=\{1,2\}$, and let $\mathbf{a}_1 \neq \mathbf{a}_2$.

Case 1: Assume $a_{n\cal N} - b^0_{\cal N} \geq 0$ for both $n=1,2$. We have $\Upsilon(\{ \mathbf{a}^i_n\}_{n \in \cal N}) = \mathbf{a}_1$ if $a_{1\cal N} < a_{2\cal N}$. Thus, player $1$ gets the maximum payoff and the game converges. Moreover, $\Phi_1(\mathbf{a}) > \Phi_1(\mathbf{a}')$ for all $\mathbf{a}' \in {\cal A}_1$ and $\Phi_{2}(\mathbf{a}) \geq \Phi_{2}(\mathbf{a}')$ for all $\mathbf{a}' \in {\cal A}_{2}$. The inequality in the latter case is tight for $\mathbf{a} \in {\cal A}_{2} \mid a_{2\cal N} > a_{1\cal N}$. Thus, player $1$ has a strongly dominant strategy while the strategy for player $2$ is weakly dominant.

Case 2: Assume $a_{n\cal N} - b^0_{\cal N} \leq 0$ for both $n=1,2$. We have $\Upsilon(\{ \mathbf{a}^i_n\}_{n \in \cal N}) = \mathbf{a}_1$ if $a_{1\cal N} > a_{2\cal N}$. Thus, player $1$ gets the maximum payoff and the game converges. Similarly, $\Phi_1(\mathbf{a}) > \Phi_1(\mathbf{a}')$ for all $\mathbf{a}' \in {\cal A}_1$ and $\Phi_{2}(\mathbf{a}) \geq \Phi_{2}(\mathbf{a}')$ for all $\mathbf{a}' \in {\cal A}_{2}$. The inequality in the latter case is tight for $\mathbf{a} \in {\cal A}_{2} \mid a_{2\cal N} < a_{1\cal N}$. Therefore, player $1$ has a strongly dominant strategy while the strategy for player $2$ is weakly dominant.

If $\mathbf{a}_1 = \mathbf{a}_{2}$, both players get a maximum payoff. The game converges. The strategy of both players is strongly dominant.
\end{IEEEproof}

For $N \geq 3$, we propose an alternative game, with faster convergence, although towards a suboptimal point. Here, the game is played between the players in a subset $S\subseteq\cal{N}$, and the dimension is reduced by only negotiating the parameters $b_{\{n\}}, n \in \cal S$ and $b_{\cal S}$. Now, the strategy of a player becomes
\be
\begin{aligned}
\mathbf{a}_n = & \hspace{4mm} \text{arg } \underset{\mathbf{a}} {\text{max}}
& & g_n(\mathbf{a}) \\
& \hspace{4mm} \text{subject to}
& & \mathbf{a} \in {\cal A}_{n\cal S}
\end{aligned}
\label{N_P_Str_ODSG}
\ee
where
\be 
{\cal A}_{n \cal S} = \{\mathbf{a} \in {\cal A}_n \mid a_{\cal T} = b^0_{\cal T} \text{ for all } {\cal T} \in {\cal P}_n, {\cal T} \neq {\cal S}, |{\cal T} | > 1 \}.
\label{Str_S_ODSG}
\ee
However, the players need to have an $\textit{a priori}$ agreed rule on how to choose the sequence of the subsets. With this approach, the players may play a sequence of single dimensional games as summarized in Algorithm~\ref{alg:2}.
\begin{algorithm}[h]
\caption{Sequential single-dimensional subset game}
\begin{algorithmic}[1]
\STATE INITIALIZATION
\STATE Given $\mathbf{b}^0$
\REPEAT
\STATE ${\cal P}' = {\cal P}$ where $\cal P$ is the power set of ${\cal N}$
\WHILE {${\cal P}' \neq \emptyset$}
\STATE Agree on a subset ${\cal S} \in {\cal P}'$
\STATE Each player $n \in \cal N$ evaluates ${\cal A}_{n\cal S}$ using~\eqref{Str_S_ODSG}.
\STATE Each player $n \in \cal N$ finds $\mathbf{a}_n \in {\cal A}_n$ using~\eqref{N_P_Str_ODSG}.
\STATE $\mathbf{b}^0 \leftarrow \Upsilon(\{ \mathbf{a}_n\}_{n \in \cal N})$
\STATE ${\cal P}' \leftarrow {\cal P}' \setminus {\cal S}$
\ENDWHILE
\UNTIL {\text convergence}
\end{algorithmic}
\label{alg:2}
\end{algorithm}

\begin{mythe}
The sequential single dimensional $N$-person game converges. 
\end{mythe}
\begin{IEEEproof}
Define $\Phi(\mathbf{a}) = \sum_{n \in \cal N} \Phi_n(\mathbf{a})$. The utility of a player $g_n(\mathbf{a})$ is concave along the line $\mathbf{b}^0+t(\mathbf{a}_n-\mathbf{b}^0), t \in [0,1]$ with the optimal value obtained at $t = 1$. Thus, for any outcome of the resolution rule and iteration $i$, we have $\Phi_n(\mathbf{a}^i) \geq \Phi_n(\mathbf{a}^{i-1})$ for all $n \in \cal N$. If $\mathbf{b}^{0^{i-1}} \neq \Upsilon(\mathbf{a}^i)$, there are at least two players such that the inequality is not tight. Therefore, we have $\Phi_n(\mathbf{a}^i) > \Phi_n(\mathbf{a}^{i-1})$ as long as $\mathbf{b}^{0(i-1)} \neq \Upsilon(\mathbf{a}^i)$. The sequential $N$-person game must converge since $\Phi(\mathbf{a})$ is bounded from above.
\end{IEEEproof}

Note that if the players play only a single dimensional sequential game along $b_n, n \in \cal N$ and $b_{\cal N}$, the strategy of the players becomes dominant with at least one of the players having a strongly dominant strategy. This can be proven using the same technique as the proof for lemma~\ref{lem:DomStr}.
\section{Applications: inter-operator resource sharing}
\label{sec:AppInt}
The $N$-person game can be applied for inter-operator resource sharing. A player would now become an operator which typically serves multiple users. Let player $n \in {\cal N}$ has multiple transmitters denoted by a set ${\cal V}_n$. Each transmitter $v \in {\cal V}_n$ may serve multiple users given by ${\cal U}_{nv}$. We have ${\cal U}_n = \cup_{v \in {\cal V}_n}{\cal U}_{nv}$. For simplicity, we assume that each transmitter and user has a single antenna.

A player allocates the resources it has to its user. We can assume the resource of a player is infinitely divisible from practicality perspective. Thus for each
${\cal S}\ni n$ and each $u\in {\cal U}_n$ player $n$ gives a fraction
$w_{u{\cal S}}$ of $b_{\cal S}$ to $u$. The user now has ``rate''
\be
 r_u = \sum_{{\cal S}\in{\cal P}_n} w_{u{\cal S}} \mu_{u{\cal S}} 
\ee
where $\mu_{u{\cal S}}$ is the ``spectral efficiency'' of user $u$ in the resource that is shared by the players in the subset $\cal S$. The spectral efficiency of a user is given as
\be
\mu_{u \cal S} = \log_2(1+\gamma_{u \cal S}),
\label{eq:SpecEff}
\ee
where
\be
\gamma_{u \cal S} = \frac{
P_{v} \lvert h_{vu} \rvert^2} 
{\underbrace{ \hspace{0mm} \sum_{ \hspace{-1mm} v' \in {\cal V}_n, v' \neq v } \hspace{0mm} 
{P_{v'} \lvert h_{v'u} \rvert^2}}_{ \text{Intra-operator interference} }\hspace{0mm} + 
\hspace{-2mm}\underbrace{\hspace{-1.0mm} \sum_{ \hspace{-1.0mm} n' \in \mathcal{S}, n' \neq n, v' \in \mathcal{V}_{n'}} \hspace{-2.0mm} {P_{v'} \lvert h_{v'u} \rvert^2}}_{ \text{Inter-operator interference}  } \hspace{-0mm} + 
\underbrace{\sigma^2}_{\text{Noise}}}
\label{eq:SINR}
\ee
is its Signal-to-noise plus interference (SINR) ratio. The transmit power budget per Hz of transmitter $v$ and the noise power per Hz on user $k$ are $P_{v}$ and $\sigma^2$, respectively. For simplicity, we assume the transmit power is uniformly distributed across the spectrum resource. The channel is given as $h_{vu} = \widetilde{h}_{vu}  / \sqrt{L_{vu}}$ where $L_{vu}$ is the distance dependent pathloss attenuation and $\widetilde{h}_{vu}$ is the complex fast fading components of the channel.

The utility of a player is given as the sum of the utility of its users. The utility of a player depends on the resource that are allocated to the user. A player allocated its resource to its users such that its utility is maximized
\be
\begin{aligned}
g_n(\mathbf{b}) = & \hspace{1mm} \underset{\mathbf{W}}{\sup }
& &   \sum_{u\in {\cal U}_n} f(r_u)\\
& \text{subject to}
& & r_u = \sum_{{\cal S}\in{\cal P}_n} w_{u{\cal S}} \mu_{u{\cal S}}, \forall u\in {\cal U}_n \\
&&& \sum_{u\in{\cal U}_{nv}} w_{u{\cal S}} = b_{\cal S},  \quad \forall v\in{\cal V}_{n}, {\cal S} \in {\cal P}_n \\
&&& \mathbf{W} \succeq 0
\end{aligned},
\label{eq:sum utility 2}
\ee
where where $f(r)$ is a suitable concave utility function, e.g. $\alpha$-fair and $\mathbf{W}$ is a $U_n \times |{\cal P}_n|$ resource allocation matrix. We assume each player independently chooses the parameter $\alpha$ for its utility function.

\begin{mypro}
The function $g_n(\mathbf{b})$ is concave in $\mathbf{b}$ if player $n$ applies an $\alpha$-PF scheduling algorithm to optimize $\mathbf{W}$.
\label{pro:CocUti}
\end{mypro}
\begin{IEEEproof}
Let Dom($g_n$) implies the domain of $g_n$ in $(\mathbf{b},\mathbf{W})$. Define a function $h_n(\mathbf{b},\mathbf{W})$ in $(\mathbf{b},\mathbf{W})$ as follows
\begin{equation}
h_n(\mathbf{b},\mathbf{W}) = \left\{
  \begin{array}{l l}
    \sum_{u\in {\cal U}_n} f(r_u) & \quad \text{if }(\mathbf{b},\mathbf{W}) \in \text{Dom}(g_n)\\
    -\inf & \quad \text{otherwise}
  \end{array} \right.,
\end{equation}
Let ${\cal W}_n$ denotes the set of points $\mathbf{W}$ satisfying the constraints of~\eqref{eq:sum utility 2}. It is closed and convex. The domain of $g_n(\mathbf{b})$ can be expressed as $\text{Dom}(g_n(\mathbf{b})) = \{\mathbf{b} \mid (\mathbf{b},\mathbf{W}) \in \text{Dom}(g_n)\}$ for some $\mathbf{W} \in {\cal W}_n$.

Due to the use of an $\alpha$-PF scheduling algorithm, $h_n(\mathbf{b},\mathbf{W})$ is jointly concave function in $(\mathbf{b},\mathbf{W})$. Let us apply Jensen's inequality on points $\mathbf{b}_1, \mathbf{b}_2 \in \text{Dom}(g_n)$, e.g. as in~\cite[p. 88]{boyd09}. For $\epsilon > 0$, there are some $\mathbf{W}_1,\mathbf{W}_2 \in {\cal W}_n$ such that $h_n(\mathbf{b}_1,\mathbf{W}_1) \geq g_n(\mathbf{b}_1)- \epsilon$ and $h_n(\mathbf{b}_2,\mathbf{W}_2) \geq g_n(\mathbf{b}_2)- \epsilon$
Taking $\theta \in [0,1]$, we have

\begin{align*}
g_n(\theta \mathbf{b}_1 + (1-\theta) \mathbf{b}_2) &= \sup_{\mathbf{W} \in {\cal W}_n} h_n(\theta \mathbf{b}_1 + (1-\theta) \mathbf{b}_2,\mathbf{W})  \\
&\geq h_n(\theta \mathbf{b}_1 + (1-\theta) \mathbf{b}_2,\theta \mathbf{W}_1 + (1-\theta) \mathbf{W}_2) \\
&\geq \theta h_n(\mathbf{b}_1, \mathbf{W}_1) + (1-\theta) h_n(\mathbf{b}_2, \mathbf{W}_2) \\
&\geq \theta g_n(\mathbf{b}_1) + (1-\theta) g_n(\mathbf{b}_2) - \epsilon.
\label{eq:4.3}
\end{align*}
This holds for any $\epsilon > 0$. Therefore, we have
\begin{equation}
g_n(\theta \mathbf{b}_1 + (1-\theta) \mathbf{b}_2) \geq \theta g_n(\mathbf{b}_1) + (1-\theta) g_n( \mathbf{b}_2),
\end{equation}
which completes the proof\footnote{This holds also if a player applies multi-point cooperative transmission to serve its users. The proof for this case is discussed in Appendix~B.}.
\end{IEEEproof}

Proposition~\ref{pro:CocUti} indicates that the utility of the player is concave in the resource utilization pattern $\mathbf{b}$. Note that the concavity is strict if $\alpha > 0$ and $g_n$ is differentiable. Therefore, the $N$-person game can be applied for inter-operator resource sharing if the operators apply an $\alpha$-fair scheduler. The operators do not need to apply the same parameter $\alpha$ for their schedulers.

\section{Numerical results}
\label{sec:NumRes}
\begin{figure}[hbtp]
\centering
\includegraphics[scale=0.3]{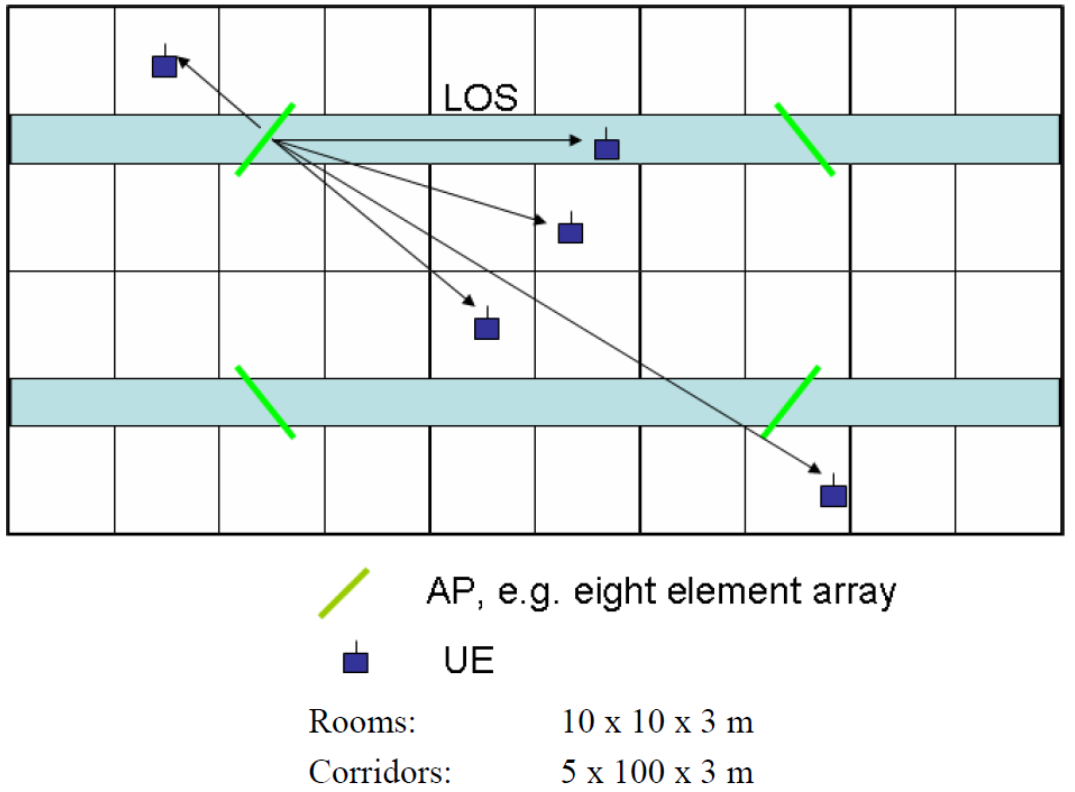}
\caption{WINNER-II A2 office layout}
\label{WINNERIIA2Layout}
\end{figure}

The performance of MRG and RPG is evaluated using inter-operator resource sharing. The operators serve users that are located in indoor office/residential environment. The layout and channel models are applied from WINNER-II model for scenario A1, i.e. indoor office~\cite{WINNERII}, see Figure~\ref{WINNERIIA2Layout}. We consider only the distance dependent pathloss and the fast fading component of the channel. There are two assumptions for the wall loss in scenario A1. We take the wall loss values for thick wall.

We consider two-and four-players games. A one unit of resource that is equal to $N\times20$ MHz is used where $N$ is the number of operators. A transmit power intensity of -53 dBm per Hz is applied which is equivalent to a power budget of 20 dBm per 20 MHz of band. The thermal noise power intensity is -195 dB which is equivalent to -121 dB per 20 MHz of band. The noise power is in general negligible when compared to intra-and inter-operator interference. 

The operators consider only the users which are served by the Transmitters (TXs) that are located in the same floor as the inter-floor interference can be considered to be small due to floor loss. Any interference from TXs that are not located in the same floor is accounted as background interference. The players may play parallel games considering the users which are associated with TXs located in each floor. 

The utility of a player is the sum of the utilities of her users which are served by the TXs located in the second floor in a three-floors building. The players are assumed to apply proportional fair scheduler, i.e. $\alpha = 1$, when allocating resources to their users. The TXs with coordinates (25,12.5), (25,-12.5), (-25,-12.5), (-25,12.5) are labelled as TXs 1, 2, 3, and 4, respectively, assuming the center of the building has a coordinate (0,0).

As a baseline, the results with default resource utilization (labelled in the Figures as 'Default') are included. As an upper bound, the results with centralized scheduler are included~\cite{Jorswieck14}, see Appendix~C. Here, we assume the cooperation between the operators is only at the Medium Access Control (MAC) layer level and there is no cooperation at the physical layer. If the constraint for the centralized scheduler include the constraint for instantaneous reciprocity, it is denoted as 'CS-SR' in the figures. If the constraint is the one for long term reciprocity, the label is 'CS-LR'. 
\subsection{Two-players game}
Let TXs 1 and 3 belong to player 1 and TXs 2 and 4 to player 2. The number of users per TX is generated using Poisson distribution with mean 5. If the probability of a user visiting the other operator's TX is 0.5, the location of the users is randomly generated with the whole floor. If the visiting probability is zero, the location of the users is distributed in a 50x25 rectangle within the floor where the own operator's TX is at the center of the rectangle. The simulation results are averaged over 100 user number and location realizations each with 20 fast fading realizations.

\begin{figure}[hbtp]
\centering
\includegraphics[scale=0.25]{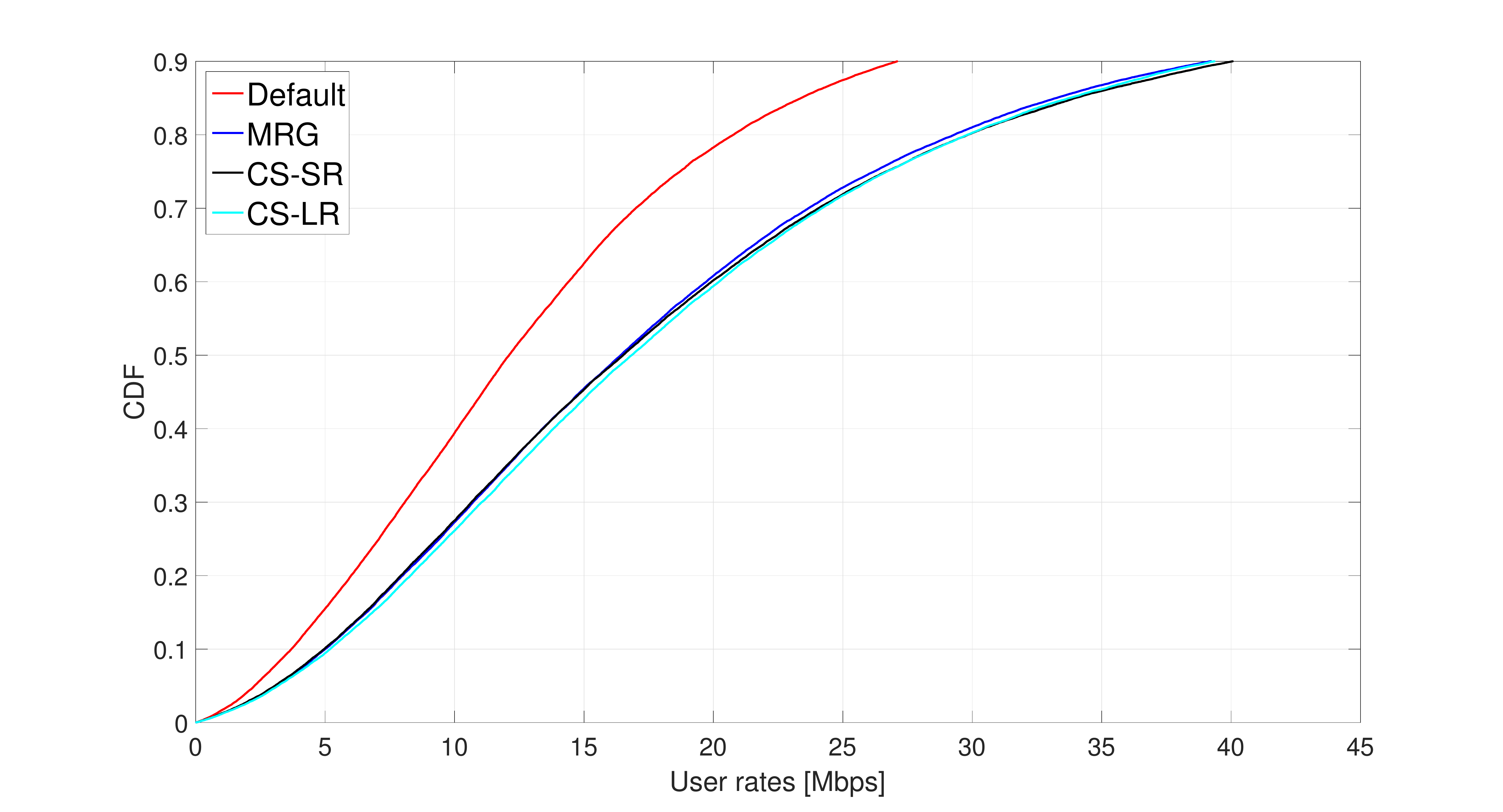}
\caption{Two-players MRG. Visiting probability = 0.}
\label{fig:twoOprsMRG_0p}
\end{figure}

\begin{figure}[hbtp]
\centering
\includegraphics[scale=0.25]{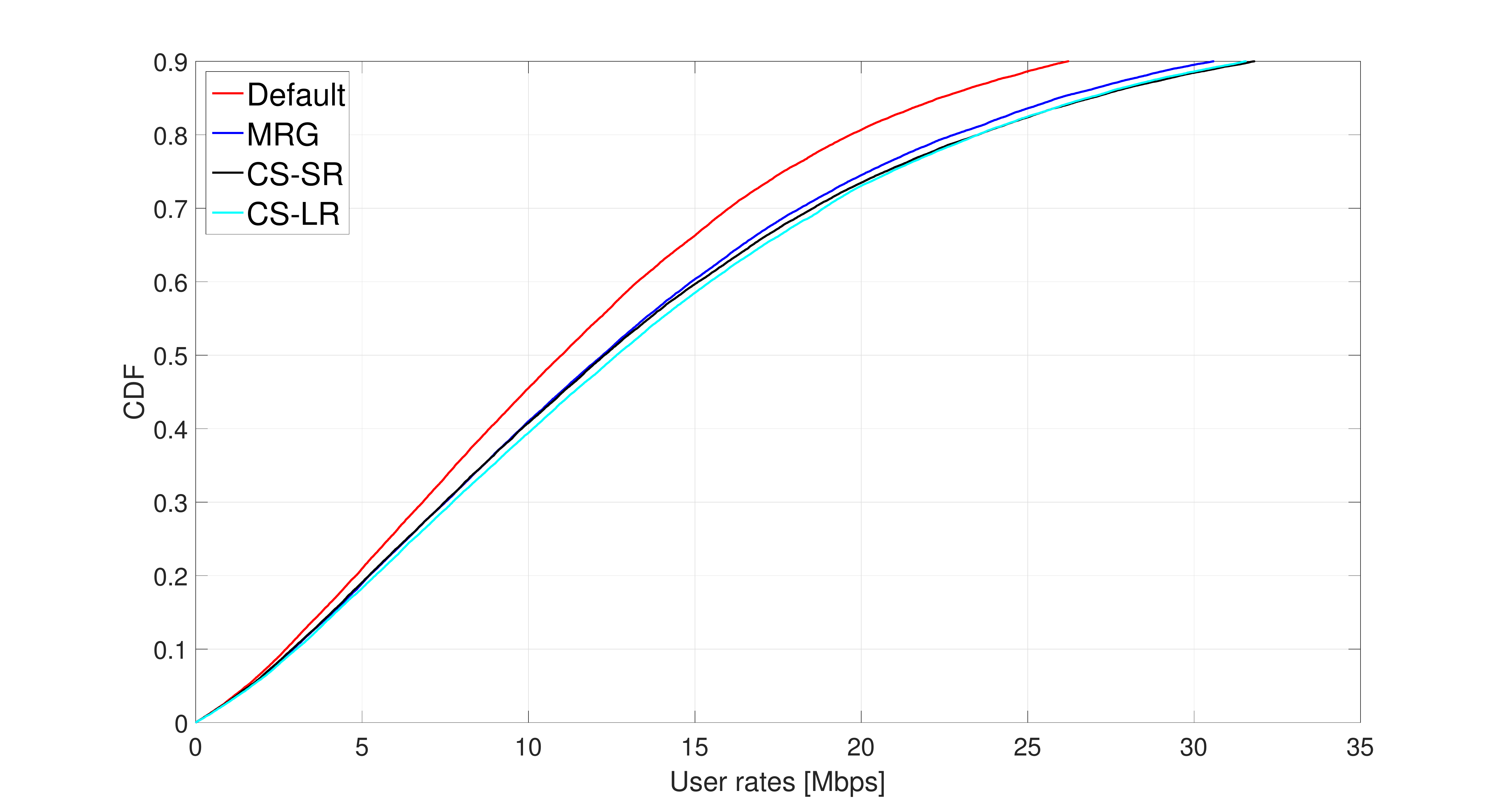}
\caption{Two-players MRG. Visiting probability = 0.5.}
\label{fig:twoOprsMRG_50p}
\end{figure}

\begin{figure}[hbtp]
\centering
\captionsetup{justification=centering}
\includegraphics[scale=0.25]{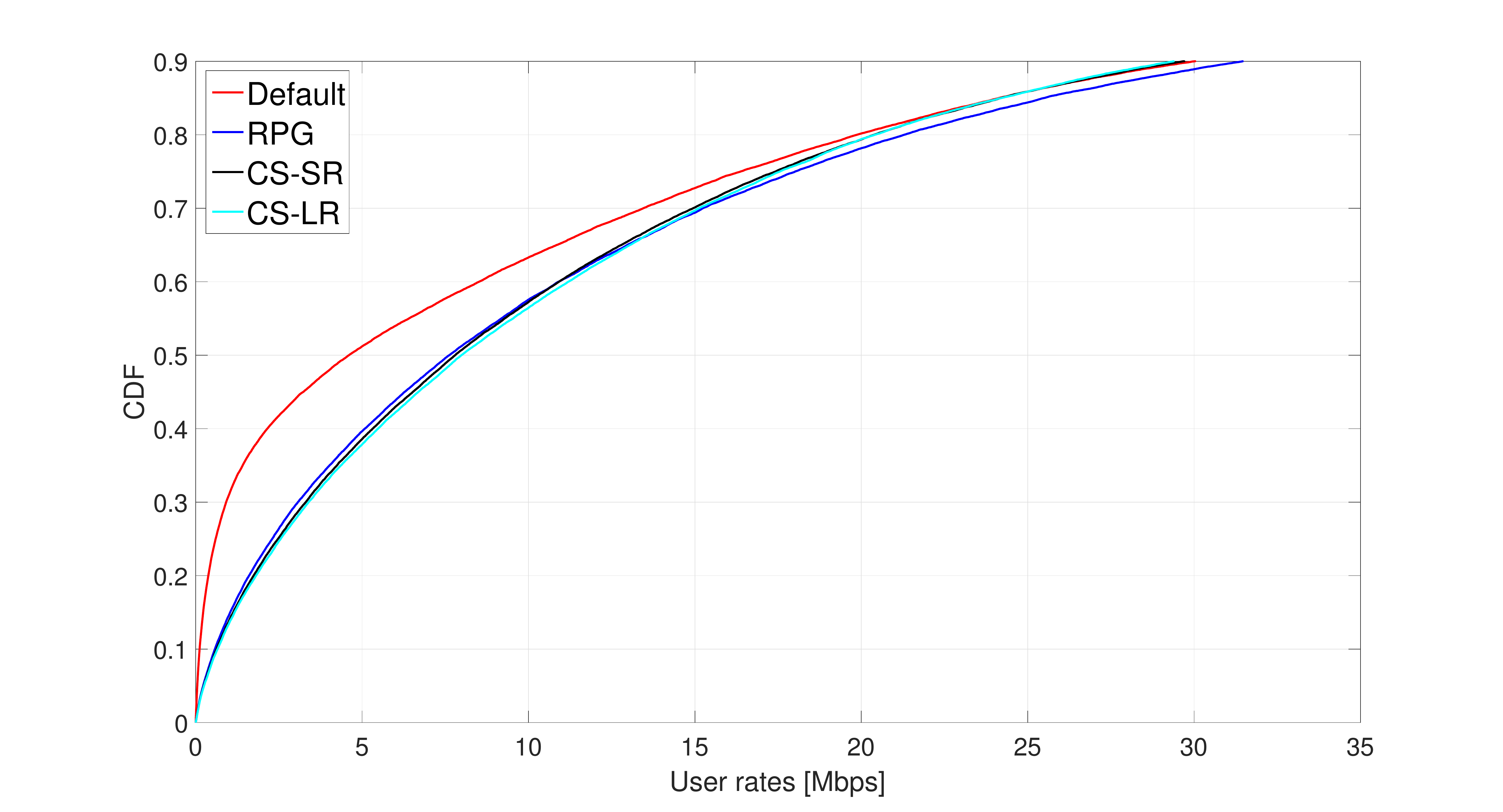}
\caption{Two-players RPG. Visiting probability = 0.5.}
\label{fig:twoOprsRPG_50p}
\end{figure}

\begin{figure}[hbtp]
\centering
\includegraphics[scale=0.25]{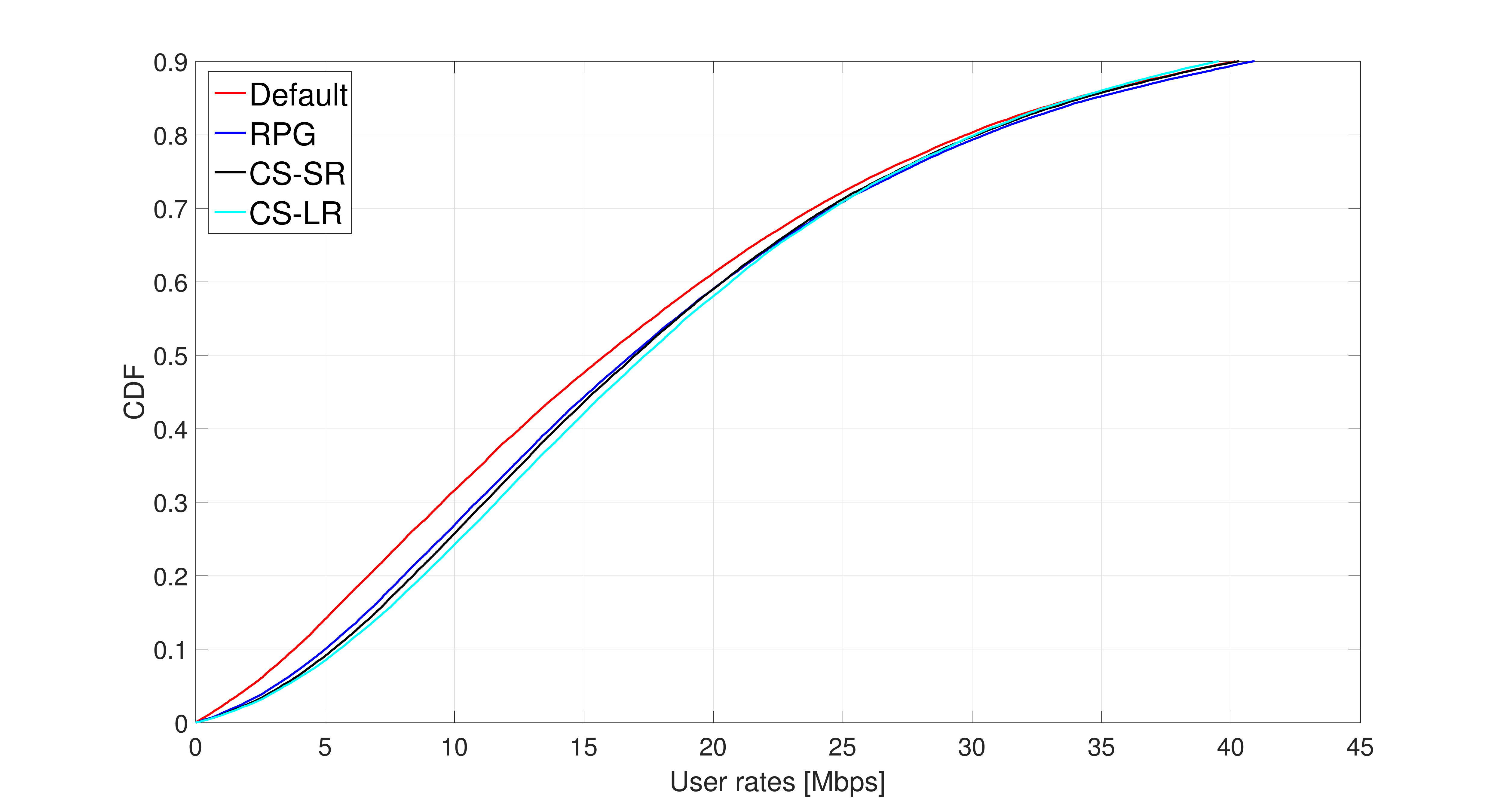}
\caption{Two-players RPG. Visiting probability = 0.}
\label{fig:twoOprsRPG_0p}
\end{figure}

Figures~\ref{fig:twoOprsMRG_0p} and~\ref{fig:twoOprsMRG_50p} show simulation results for two-players MRG and figures~\ref{fig:twoOprsRPG_50p} and~\ref{fig:twoOprsRPG_0p} for two-players RPG. The results for MRG with visiting probability of 0 and RPG with visiting probability of 0.5 have a significant gain comparing to the results with 'default' utilization pattern. The difference from the result for 'CS-SR' is almost negligible. The results for 'CS-LR' has a slightly better performance than the results for MRG (or RPG depending on the game) and 'CS-SR'. This is due to the fact that the 'CS-LR' scheduler utilizes the load difference between the players in addition to the location of the users.

The results for MRG with visiting probability of 0.5 and RPG with visiting probability of 0 are almost the same as the results for the 'default', 'CS-SR' and 'CS-LR'. The reason for this is that keeping resources orthogonalized is almost optimal when there is strong inter-operator interference (comparing to intra-operator interference plus background noise/interference) and re-using resources is close to optimal when there is negligible inter-operator interference. Due to this, the operators may consider the users which has small inter-operator interference in the MRG case and strong inter-operator interference in RPG.
\subsection{Four-players game}
Let TX $n$ belongs to player $n$ where $n \in \{1,2,3,4\}$. The number of users per TX is generated using Poisson distribution with mean 5. The location of the users is generated in a similar manner as in the two-players game. The simulation results are averaged over 50 user number and location realizations each with 10 fast fading realizations.

\begin{figure}[hbtp]
\centering
\includegraphics[scale=0.35]{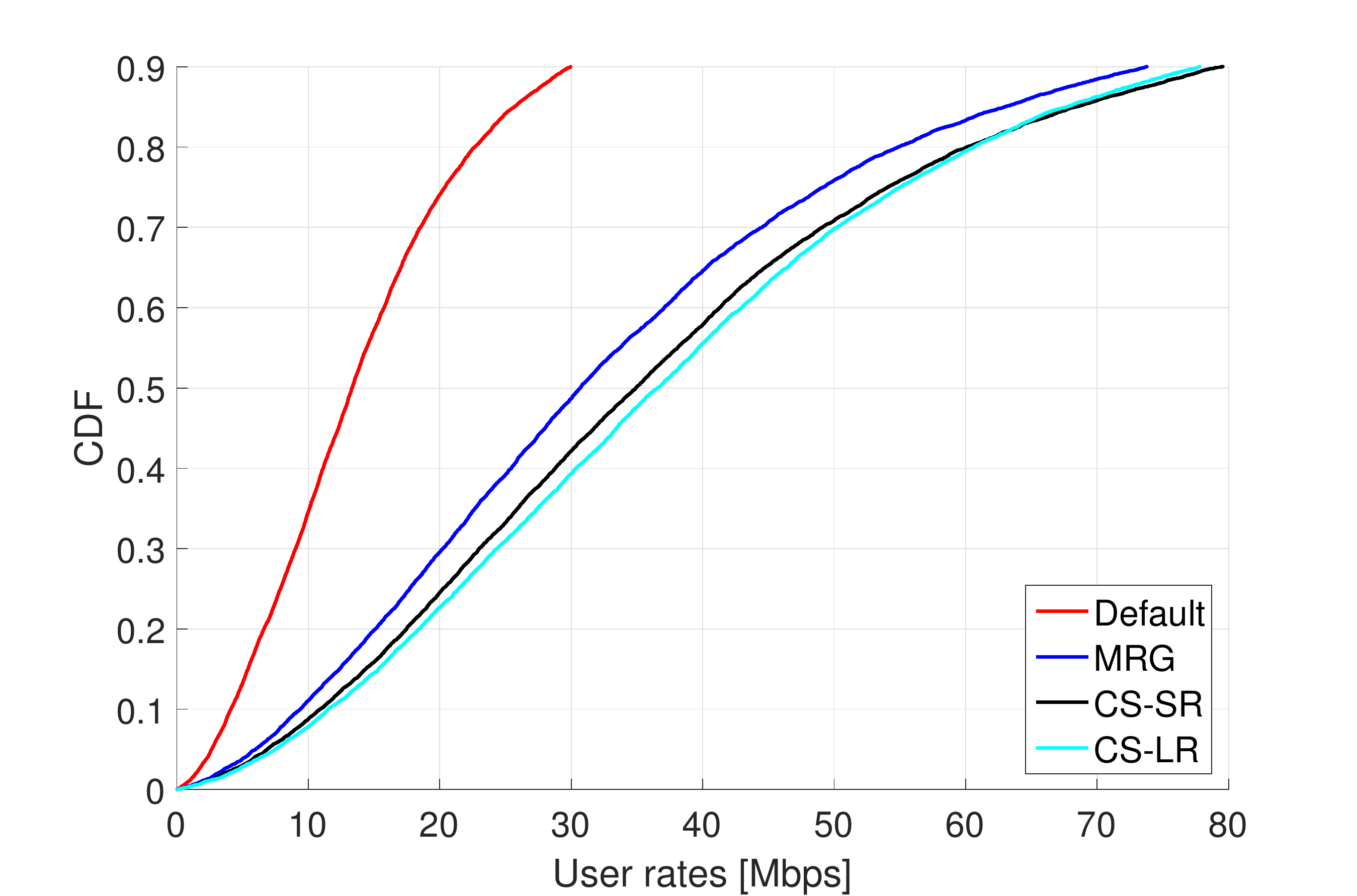}
\caption{Four-players MRG. Visiting probability = 0.}
\label{fig:fourOprsMRG}
\end{figure}

\begin{figure}[hbtp]
\centering
\includegraphics[scale=0.35]{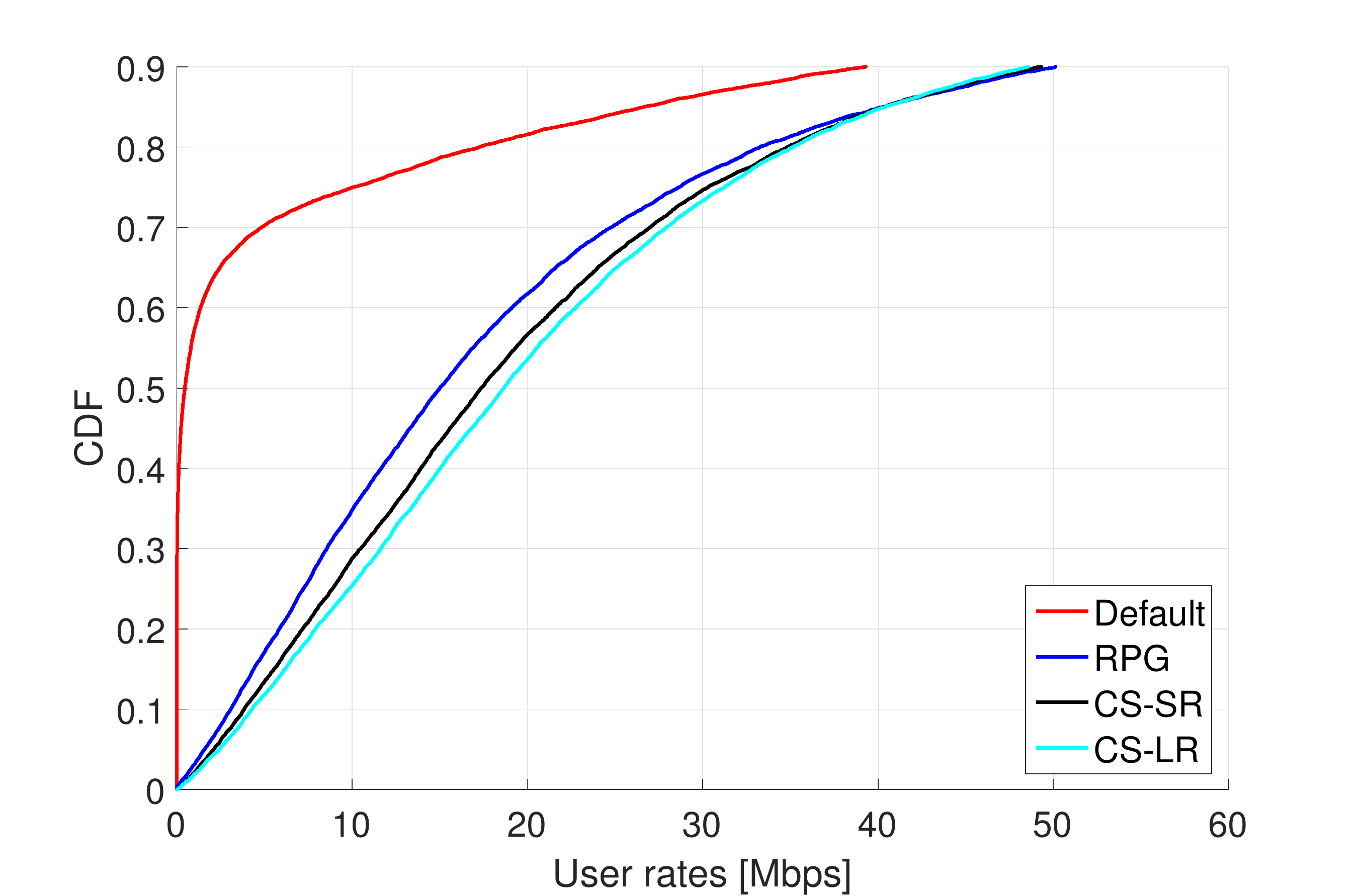}
\caption{Four-players RPG. Visiting probability = 0.5.}
\label{fig:fourOprsRPG}
\end{figure}

Figures~\ref{fig:fourOprsMRG} and~\ref{fig:fourOprsRPG} show simulation results for four-players MRG and RPG. The results for MRG and RPG are obtained such that the players first play the sequential multi-dimensional $N$-person game according Algorithm~\ref{alg:SeqNpe}. After it converges, they play sequential single-dimensional $N$-person game according to Algorithm~\ref{alg:2}. To agree on a direction in the single-dimensional game, the players first propose their preferred subsets. The subsets are given probability weights that are proportional to the number of players that voted the subset. One of the subsets is selected randomly with the probability weights. A player chooses a subsets that can ideally lead to the maximum utility improvement. The same baseline and upper bounds as the two-players game are also used for the four-players games.

The results for MRG with visiting probability of 0 and RPG with visiting probability of 0.5 have a significant gain comparing to the results with 'default' utilization pattern. However, comparison with the upper bounds indicates that there is still room for improvement.

\begin{figure}[hbtp]
\centering
\includegraphics[scale=0.3]{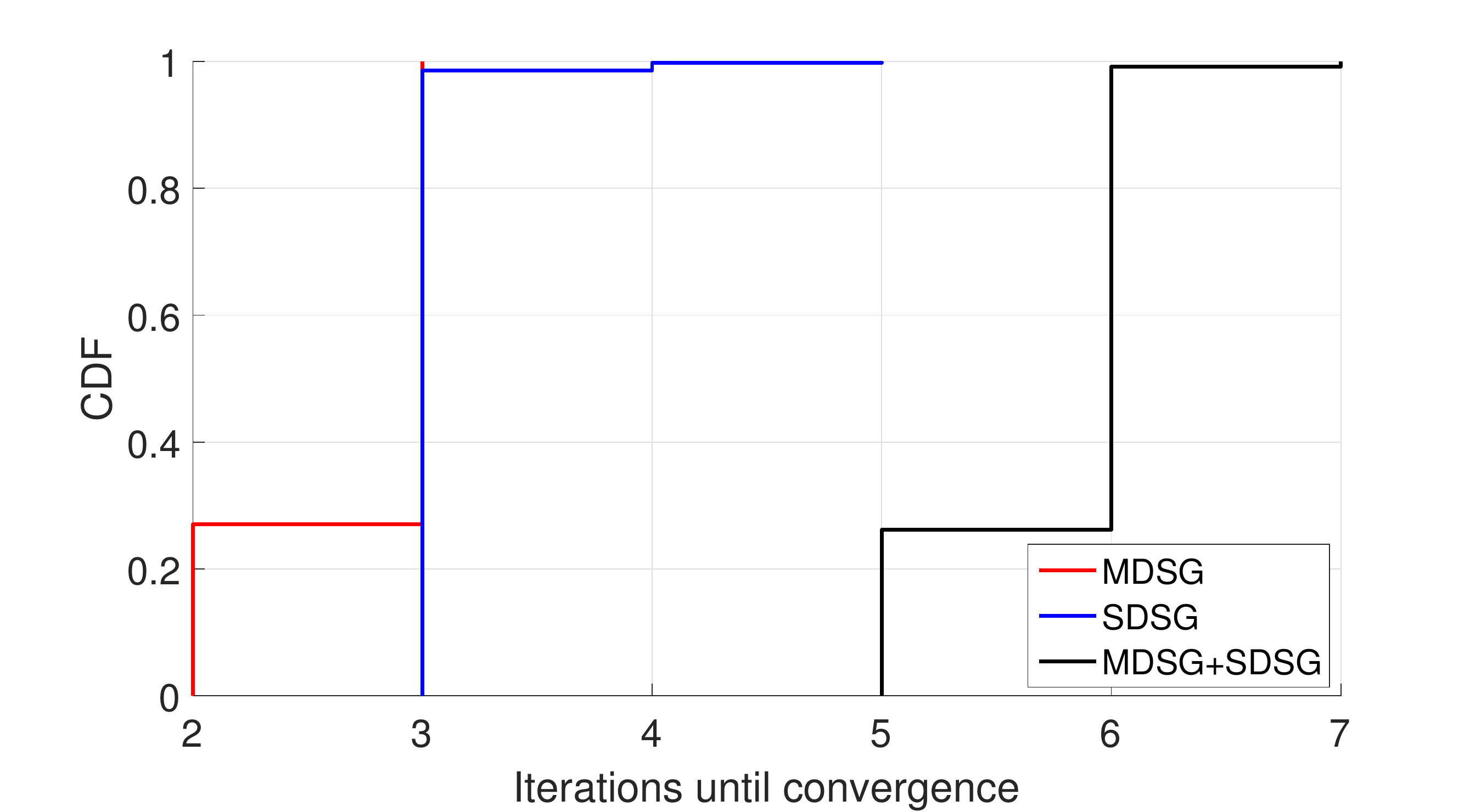}
\caption{Four-players MRG. Visiting probability = 0.}
\label{fig:convFourOprsMRG}
\end{figure}

\begin{figure}[hbtp]
\centering
\includegraphics[scale=0.3]{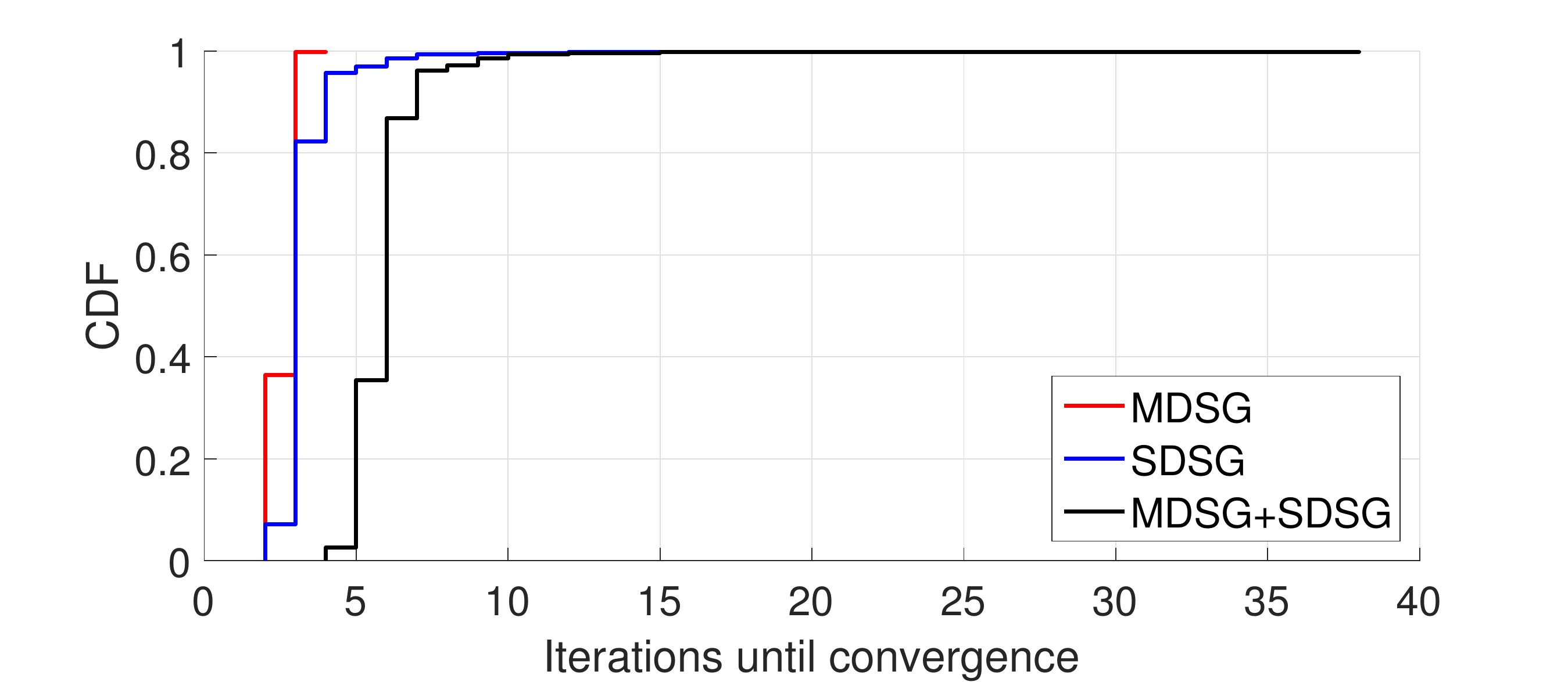}
\caption{Four-players RPG. Visiting probability = 0.5.}
\label{fig:convFourOprsRPG}
\end{figure}

On the other hand, Figures~\ref{fig:convFourOprsMRG} and~\ref{fig:convFourOprsRPG} show the number of iteration until the four-players MRG and RPG converge. The convergenceforthe sequential game summarized in Algorithm~\ref{alg:SeqNpe} is referred as 'MDSG' and the one in in Algorithm~\ref{alg:2} as 'SDSG'. The results indicate that the convergence of both the MRG and RPG games takes few iterations in most cases. 

\section{Conclusion}
\label{sec:Con}

In this paper, we have proposed a game for resource allocation between operators, who are allowed to form several coalitions simultaneously. The game was proven to have a Nash equilibrium, via geometric methods. When the game is played sequentially between greedy players, it was shown that the outcome converges, and even converges fast in a given example setting. Moreover, in our simulated setting, the outcome appears to be close to a Pareto-optimal distribution, as selected by a centralized coordinator. Further research is needed to prove (near) Pareto-optimality as well as to study the game with transparent and transferable utilities.

\section*{Appendix A: Proof of Theorem~\ref{thm:linear}}
\begin{IEEEproof}
The set \[\tilde{\mathcal{F}}=\mathcal{F}\cap(\cap_{n\in \mathcal{N}}I_n),\] over which the maximal is taken, is the intersection of a convex polytope $\mathcal{F}$ with convex polyhedra. Hence, it is a convex polytope in its own right. To prove that the objective function \[\sum_{S \subset {\cal N}} |\mathbf{b}_S - \mathbf{b}^0_S|\] is linear on the feasible set, we rewrite it as \[\sum_{S \subset {\cal N}} \alpha_S(\mathbf{b}_S - \mathbf{b}^0_S),\] where \[\alpha_S=\left\{
\begin{split} &1 &\hbox{ if } \mathbf{a}_{nS}>\mathbf{b}^0_S \hbox{ for all } n\in S \\
& -1 &\hbox{ if } \mathbf{a}_{nS}>\mathbf{b}^0_S \hbox{ for all } n\in S \\
& 0 & \hbox{ otherwise }.
\end{split}
\right.\]
By the definition of $I_n$, we see that if $\mathbf{a}_{nS} - \mathbf{b}^0_S$ has different signs for different $n\in S$, then $b_S=b_S^0$ is the only feasible value for $b_S$, so \[|\mathbf{b}_S - \mathbf{b}^0_S|=0\] for all $\mathbf{b}\in \tilde{\mathcal{F}}$ On the other hand, if for some $n\in S$ we have $\mathbf{a}_{nS} > \mathbf{b}^0_S$, then all feasible $\mathbf{b}$ will have $\mathbf{b}_S\geq\mathbf{b}^0_S$, so \[|\mathbf{b}_S - \mathbf{b}^0_S|=1\cdot(\mathbf{b}_S - \mathbf{b}^0_S)\] on $\tilde{\mathcal{F}}$. Analogously, if $\mathbf{a}_{nS} < \mathbf{b}^0_S$, then \[|\mathbf{b}_S - \mathbf{b}^0_S|=-1\cdot(\mathbf{b}_S - \mathbf{b}^0_S)\] on $\tilde{\mathcal{F}}$ Hence we have \[\sum_{S \subset {\cal N}} |\mathbf{b}_S - \mathbf{b}^0_S| = \sum_{S \subset {\cal N}} \alpha_S(\mathbf{b}_S - \mathbf{b}^0_S)\] on $\tilde{\mathcal{F}}$. This proves the theorem.
\end{IEEEproof}
\label{sec:AppA}

\section*{Appendix B: CoMP scheduler}
\label{sec:AppB}
In Section~\ref{sec:AppInt}, it is shown that the utility of the operators is concave function in the resource utilization pattern assuming there is no cooperative transmission among the transmitters. In this appendix, we show that the result hold also if the transmitters of a player apply cooperative multi-point transmission (CoMP). 

Assume the transmitters of an operator have a centralized scheduler. The transmitters can cooperatively serve the users in $C_n$ ways where $C_n = \sum_{k = 1}^{\min(U_n,V_n)} U_n!/n!(U_n-n)!$. Recall from Section~\ref{sec:AppInt} that $U_n$ denotes the number of users of operator $n$ and $V_n$ denotes its number of transmitters. 

With this setting, the spectral efficiency of a user depends on the set of users with whom it is scheduler. The inter-operator interference is, however, colored unless the operator apply a unitary precoder, see e.g.~\cite{Hailu2014}. If the interference is colored, the operator might be assumed to estimate the inter-operator interference based on expected values. With this assumption, let $\mu_{uc{\cal S}}$ denote the spectral efficiency of user $u$ when the transmitters cooperatively serve the users in user group ${\cal C}_c \subseteq {\cal U}_n$ where $c \in \{1,\hdots,C_n\}$. Note that $\mu_{uc{\cal S}} = 0$ if user $u$ is not a member of the user group ${\cal C}_c$.

Thus, the rate of a user becomes
\be
 r_u = \sum_{{\cal S}\in{\cal P}_n} \sum_{c=1}^{C_n} w_{uc{\cal S}} \mu_{uc{\cal S}} 
\ee
where $w_{uc{\cal S}}$ the resource allocated to user group $c$ from the resource $b_{\cal S}$. Now, the resource allocation matrix $W$ is a $U_n \times C_n \times |{\cal P}_n|$ matrix. The utility of a player becomes
\be
\begin{aligned}
g_n(\mathbf{b}) = & \hspace{1mm} \underset{\mathbf{W}}{\sup }
& &   \sum_{u\in {\cal U}_n} f(r_u)\\
& \text{subject to}
& & r_u = \sum_{{\cal S}\in{\cal P}_n} \sum_{c=1}^{C_n} w_{uc{\cal S}} \mu_{uc{\cal S}} \\
&&& \sum_{c=1}^{C_n} w_{uc{\cal S}} = b_{\cal S}, \quad {\cal S} \in {\cal P}_n \\
&&& \mathbf{W} \succeq 0
\end{aligned}.
\label{eq:sum utility comp}
\ee
Observe that the utility of a player is a jointly concave function in $\mathbf{b}$ and $\mathbf{W}$ if it uses an $\alpha$-fair scheduler. The same technique as the proof for Proposition~\ref{pro:CocUti} can be used to prove that $g_n$ is a concave function in $\mathbf{b}$.

\section*{Appendix C: Centralized Scheduler}
\label{sec:AppC}
In the Section~\ref{sec:NumRes}, we included the result for centralized scheduler which determines the spectrum utilization patter for the operators such that the sum of the utilities of the players\footnote{The sum utility can be changed into a weighted sum of the utilities of the players in a straight forward manner.} is maximized. The sum utility can be maximized such that the instantaneous reciprocity is fulfilled as
\be
\begin{aligned}
\mathbf{b}^* = & \hspace{4mm}\argmax[\mathbf{b}]
& & \sum_{n \in \cal N} g_n(\mathbf{b}) \\
& \hspace{4mm} \text{subject to}
& &  \sum_{S\in{\cal P}_n} \frac{b_S}{|S|} = \frac1N \quad \forall n \in {\cal N} \\
&&& \mathbf{b} \succeq 0.
\end{aligned}
\label{eq:CS_SR}
\ee
Such a centralized scheduler might be opted by the operators, for example, if they don't have symmetric loads.

On the other hand, the operators might opt for a centralized scheduler with long-term reciprocity, i.e. the expected favors that are given and taken by each operator is equal. The resource usage pattern in this case is given as
\be
\begin{aligned}
\mathbf{b}^* = & \hspace{4mm}\argmax[\mathbf{b}]
& & \sum_{n \in \cal N} g_n(\mathbf{b}) \\
& \hspace{4mm} \text{subject to}
& &  \sum_{S \subset {\cal N}} b_S = 1~ \\
&&& \mathbf{b} \succeq 0.
\end{aligned}
\label{eq:CS_LR}
\ee
Such a scheduler is especially beneficial if the operators have a symmetric load. 

Note that the objective function of both~\eqref{eq:CS_SR} and~\eqref{eq:CS_LR} is concave function as it is a sum of concave functions. The centralized scheduler may also need to know the spectral efficiencies of the users of the operators in order to obtain the utility function of the players as a function resource utilization pattern and solve the problem.

\section*{Acknowledgment}

This work was supported in part by the European Commission in the
framework of the H2020 project ICT-671639 COHERENT.

\bibliographystyle{IEEEtran}
\bibliography{IEEEabrv,hailu16_jrnl_comsoc}

\end{document}